\documentclass[preprint2]{aastex62}

\graphicspath{{./}{figures/}}

\received{}
\revised{}
\accepted{}
\submitjournal{ApJ}

\shorttitle{HST/STIS capability for Love number measurement}
\shortauthors{Hellard et al.}

\begin{document}

\title{HST/STIS capability for Love number measurement of WASP-121b}

\correspondingauthor{Hugo Hellard}
\email{hugo.hellard@dlr.de}

\author[0000-0003-0077-3196]{Hugo Hellard}
\affiliation{Deutsches Zentrum f\"ur Luft und Raumfahrt, Institut f\"ur Planetenforschung \\
Rutherfordstra\ss e 2, 12489 Berlin,
Germany, DE}
\affiliation{Technische Universit\"at Berlin \\
Stra\ss e des 17. Juni 135, 10623
Berlin, Germany, DE}

\author{Szil\'ard Csizmadia}
\affiliation{Deutsches Zentrum f\"ur Luft und Raumfahrt, Institut f\"ur Planetenforschung \\
Rutherfordstra\ss e 2, 12489 Berlin,
Germany, DE}

\author{Sebastiano Padovan}
\affiliation{Deutsches Zentrum f\"ur Luft und Raumfahrt, Institut f\"ur Planetenforschung \\
Rutherfordstra\ss e 2, 12489 Berlin,
Germany, DE}

\author{Frank Sohl}
\affiliation{Deutsches Zentrum f\"ur Luft und Raumfahrt, Institut f\"ur Planetenforschung \\
Rutherfordstra\ss e 2, 12489 Berlin,
Germany, DE}

\author{Heike Rauer}
\affiliation{Deutsches Zentrum f\"ur Luft und Raumfahrt, Institut f\"ur Planetenforschung \\
Rutherfordstra\ss e 2, 12489 Berlin,
Germany, DE}
\affiliation{Technische Universit\"at Berlin \\
Stra\ss e des 17. Juni 135, 10623
Berlin, Germany, DE}
\affiliation{Freie Universit\"at Berlin, Institut f\"ur Geologische Wissenschaften \\
Malteserstra\ss e 74-100, 12249 Berlin,
Germany, DE}



\begin{abstract}

Data from transit light curves, radial velocity and transit timing observations can be used to probe the interiors of exoplanets beyond the mean density, by measuring the Love numbers $h_2$ and $k_2$. The first indirect estimate of $k_2$ for an exoplanet from radial velocity and transit timing variations observations has been performed by taking advantage of the years-spanning baseline. Not a single measurement of $h_2$ has been achieved from transit light curves, mostly because the photometric precision of current observing facilities is still too low. We show that the Imaging Spectrograph instrument on-board the Hubble Space Telescope could measure $h_2$ of the hot Jupiter WASP-121b if only few more observations were gathered. We show that a careful treatment of the noise and stellar limb darkening must be carried out to achieve a measurement of $h_2$. In particular, we find that the impact of the noise modelling on the estimation of $h_2$ is stronger than the impact of the limb darkening modelling. In addition, we emphasize that the wavelet method for correlated noise analysis can mask limb brightening. Finally, using presently available data, we briefly discuss the tentative measurement of $h_2 = 1.39^{+0.71}_{-0.81}$ in terms of interior structure. Additional observations would further constrain the interior of WASP-121b and possibly provide insights on the physics of inflation. The possibility of using the approach presented here with the Hubble Space Telescope provides a bridge before the high-quality data to be returned by the James Webb Space Telescope and PLATO telescope in the coming decade.

\end{abstract}

\keywords{HST photometry --- 
Planets and Satellites: interiors --- Planets and Satellites: individual (WASP-121b)}


\section{Introduction} \label{sec:intro}

Celestial bodies, and in particular planets orbiting close to their Roche limit will deform in response to tidal interactions with their host star and, to a lower extent, to their own rotation. The radial deformation of the planetary outer shape can be quantified using the Love number $h_2$, while the redistribution of mass within the planet can be expressed through the Love number $k_2$ \citep{love1911}. The perturbing potential, $V_p$, is usually expressed as a sum of harmonics of degree n, $V_{p,n}$:
\begin{equation}
    V_p = \sum_{n=2}^{\infty}V_{p,n}.
\end{equation}
For example, the tidal potential of degree n, $V_{t,n}$, is equal to:
\begin{equation}
    \label{eq:tidalpotential}
    V_{t,n} = \frac{Gm_s}{d}\Bigg(\frac{R_p}{d}\Bigg)^{n}P_n(\lambda),
\end{equation}
where $G$ is the gravitational constant, $m_s$ is the stellar (i.e., perturber) mass, $d$ is the center-to-center distance, $R_p$ is the mean planetary radius (i.e., the radius the planet would have if it were isolated and non-rotating), $P_n$ is the Legendre polynomial of degree $n$, and $\lambda = \sin\theta\cos\varphi$, with $\theta$ and $\varphi$ the colatitude and east longitude, respectively.\\
Likewise, the resulting potential at the planet $V_r$, and the total radial deformation $\Delta r$ can be expressed as:
\begin{eqnarray}
    \label{eq:vr}
    V_r & = & \sum_{n=2}^{\infty}V_{r,n} \\
    \label{eq:deltar}
    \Delta r & = & \sum_{n=2}^{\infty} \Delta r_n
\end{eqnarray}
The Love numbers of degree $n$, $h_n$ and $k_n$ are defined as \citep{love1911,kopal1959}:
\begin{eqnarray}
    \Delta r_n & = & h_n\frac{V_{p,n}}{g} \\
    V_{r,n} & = & k_n V_{p,n},
\end{eqnarray}
where $g$ is the mean surface acceleration of the unperturbed planet. The derivation of the Love numbers depend on the planetary interior. In case of hydrostatic equilibrium (i.e., the interior behaves as a fluid), $h_n$ and $k_n$ only depend on the radial density profiles, and there is a simple relation $h_n = 1 + k_n$ with $0 < k_n < 1.5$ \citep{munk1960}. A value of 0 would describe a point-mass surrounded by a mass-less envelope, while a value of 1.5 would represent a homogeneous body. For instance, $k_2 \simeq 0.985$ for the fluid Earth \citep{lambeck1980}, and $k_2 \simeq 0.6$ for Jupiter \citep{iess2018}. While hydrostatic equilibrium is a fair assumption for hot Jupiters - since gas behaves as a fluid - it may not be true for ice giants or rocky planets. In that case, the Love numbers depend on additional interior parameters such as rigidity and viscosity, and on the timescale of the perturbation \citep{correia2013}. In either case, the knowledge of these numbers provides additional information on planetary interiors \citep[e.g.,][]{baumeister2019,kellermann2018,padovan2018}. We usually focus on the second degree Love numbers $h_2$ and $k_2$ because they are the most sensitive to the interior and the easiest to measure \citep{gavrilov1977}. Therefore we have $0 < h_2 < 2.5$, where 0 represents an infinitely rigid body (i.e., a non-deformable sphere) while 2.5 describes a homogeneous fluid body. The interior of WASP-121b is expected to be in hydrostatic equilibrium since it is mainly composed of gas, and the latter behaves as a fluid. In addition, close-in giants are expected to be locked in a 1:1 spin-orbit resonance \citep{ragozzine2009}, a configuration for which the linear approximation for the computation of the fluid Love numbers holds \citep{padovan2018}. In that case the theoretical range of values for $h_2$ is between 1 and 2.5. However, non-linear effects may change the value of the Love numbers by several percents \citep{wahl2016,wahl2017}. Accordingly, we opt for a less-biased wider prior for $h_2$ which can accommodate these a priori-unknown effects. We do not extend the upper bound on $h_2$ over 2.5 because it is highly unlikely to have a gas giant close to being homogeneous, since it is mostly made of gas, which in the outer part is very compressible.\\
\\
It has been shown that $k_2$ can be indirectly measured from radial velocity variations (RVVs) and/or transit timing variations (TTVs) observations \citep{csizmadia2019}, while $h_2$ can be measured from transit light curve (TLC) observations \citep{correia2014,akinsanmi2019,hellard2019}. Depending on the assumed system's dynamics (e.g., stellar and planetary rotation rates, orbits), estimates of $k_2$ recently became available for few exoplanets \citep[e.g.,][]{buhler2016,hardy2017,bouma2019,csizmadia2019}, while $h_2$ has never been measured from TLCs. This is because the photometric uncertainty of current telescopes is still above the required levels. The upcoming JWST (launch planned for 2021) and PLATO (launch planned for 2026) missions will provide accurate enough TLC observations to precisely measure $h_2$ \citep{hellard2019}. In this paper we show that, in the meantime, the Imaging Spectrograph (STIS) instrument on-board the Hubble Space Telescope (HST) has the capability to carry out Love number measurements of WASP-121b from TLCs. We prove this point by using two transits observed by HST/STIS (Section \ref{sec:data}). In Section \ref{sec:transit_model} we present the transit model. To achieve a precise measurement of $h_2$, we show in Section \ref{sec:lc_analysis} that a careful treatment of the noise should be performed and that the wavelet method performs well for that purpose. Furthermore, a cautious modelling of the stellar limb darkening (LD) should be used, and a thorough analysis of the fitted stellar intensity profile must be carried out. We conclude with a discussion on the retrieved Love number and its potential to better understand the physics of inflation (Section \ref{sec:discussions}).

\section{Data} \label{sec:data}
\begin{figure}
    \centering
    \includegraphics[width=0.5\textwidth]{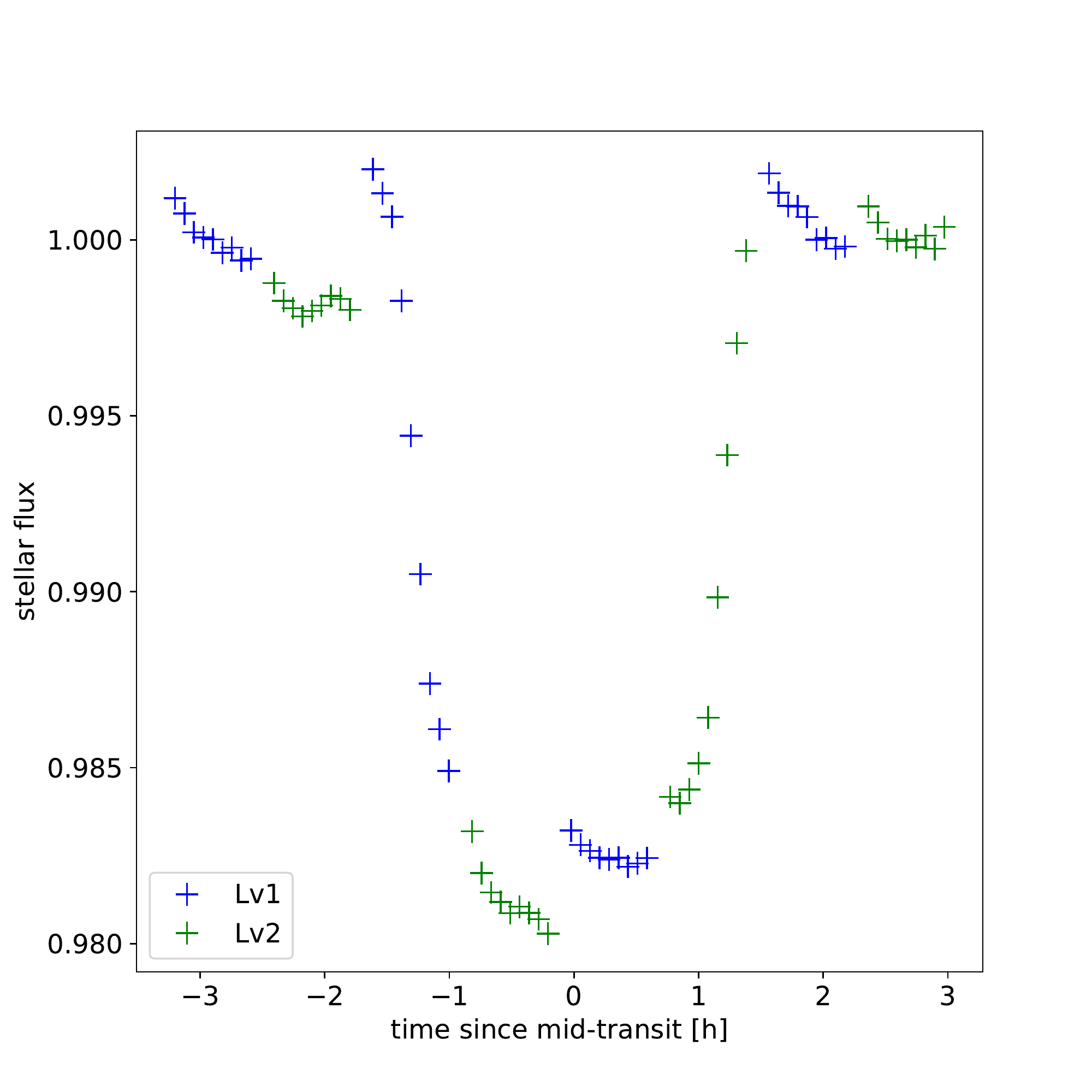}
    \caption{HST/STIS raw data for Lv1 and Lv2. The systematics due to thermal breathing are clearly visible.}
    \label{fig:raw_data}
\end{figure}

We used observations of two primary transits of WASP-121b made by HST/STIS, in the frame of the Panchromatic Comparative Exoplanet Treasury (PanCET) survey (Progam 14767; P.I.s D. Sing and M. L\'opez-Morales). The first visit was made on October 24, 2016 (hereinafter Lv1) and the second one on November, 6, 2016 (hereinafter Lv2), with the G430L grating. The data were published in \citet{evans2018}, where the reader can find further details on the data reduction process. In \citet{evans2018}, two other primary transits were published (one with the G750L grating and another with the G141 grism of the Wide Field Camera 3). However, the data of these two extra transits do not cover the ingress and egress phases of the transit, which are important for Love number measurements \citep{carter2010,correia2014}. Accordingly, we did not use these two transits in our analysis\footnote{We ran an additional analysis included these two transits and did not obtain any improvements on the $h_2$ measurement.}. The raw white light curves are presented in Figure \ref{fig:raw_data}.
One can easily identify the systematics coming from the thermal expansion and contraction of the telescope along each orbit, the so-called thermal breathing of HST\footnote{see STScI Instrument Science Report ACS 2008-03}. The treatment of such time-correlated noise is crucial when retrieving information from TLCs (see Section \ref{sec:noise_analysis}). In the next section we briefly present the transit model used to retrieve the Love number $h_2$ from the TLC.

\section{Transit model} \label{sec:transit_model}
The object WASP-121b orbits its F6V host star at a distance of roughly 0.03 AU in about 1.3 days \citep{delrez2016}. Considering the mass estimates available in the literature \citep{delrez2016}, this results in a planet orbiting at 2.13 Roche radii, where the Roche radius corresponds to the orbital distance at which tidal interactions overcome the planet's self gravity. Therefore, WASP-121b is highly deformed due to tidal interactions with its host star, and the planetary shape should not be modelled as a sphere. To further support that, we calculate a simple estimate of the radial deformation the planet experiences due to tides. Using the values from \citet{delrez2016} and assuming a Jovian-like Love number $h_2=1.6$, we derive $\Delta r_2 / R_p = 0.05$ with Equations (\ref{eq:tidalpotential}) and (\ref{eq:deltar}) . A radial deformation of 5\%, which translates into roughly 3600 km, is clearly not negligible. Therefore, we adopt the shape model described by \citet{hellard2019}, where the radius is given by:
\\
\\
\begin{eqnarray}
\label{eq:3dshape}
r(\theta,\varphi) &=& R_p\Bigg(1 + q\sum_{n=2}^{4}h_nP_n(\lambda)\left(\frac{R_p}{d}\right)^{n+1} \\
\nonumber
&-& \frac{1}{3}h_2(1+q)F_p^2\left(\frac{R_p}{d}\right)^3P_{2}(\cos(\Theta))\Bigg),
\end{eqnarray}
where $q$ is the ratio between the stellar and planetary masses, $F_p$ is the ratio between the orbital and rotational periods, $\Theta$ is the obliquity. We assume a circular orbit \citep{delrez2016} ($d$ is equal to the semi-major axis), a synchronized rotation ($F_p=1$), and a rotational axis perpendicular to the orbital plane ($\Theta=\theta$). Because the sensitivity of the planetary Love numbers to the interior decreases with increasing degree \citep[see, e.g.,][]{padovan2018}, we fix $h_3 = h_4 = 1$ and assume a spherical star.\\
\\
The integration of the occulted stellar area during transit phases result in the modelled TLC \citep{hellard2019}. The fitted transit parameters are the inclination ($i$), the epoch ($E_0$), the limb darkening coefficients (LDCs, see Section \ref{sec:ld_analysis}), the normalized semi-major axis ($d/R_s$), the normalized planetary mean radius ($R_p/R_s$), and $h_2$.

\section{Light curve analysis} \label{sec:lc_analysis}

\subsection{Noise analysis} \label{sec:noise_analysis}
\begin{figure*}
    \plotone{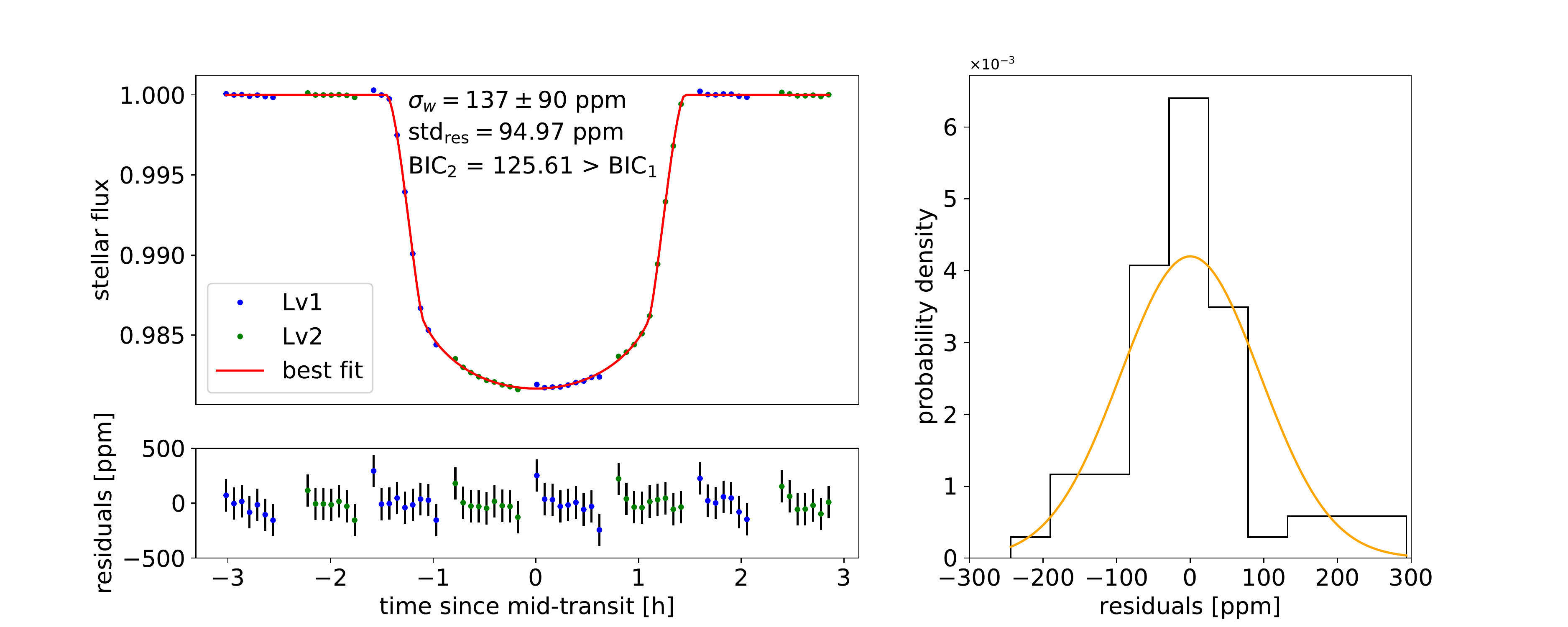}
    \caption{Systematics-free data and best fit (top left panel), residuals with white noise errorbars (bottom left panel), and distribution of the residuals (right panel) for Model 1. $\mathrm{std_{res}}$ corresponds to the standard deviation of the residuals, while the orange solid line in the right panel corresponds to the Gaussian fit of the residuals' distribution.}
    \label{fig:g430l_model1}
\end{figure*}

\begin{figure*}
    \plotone{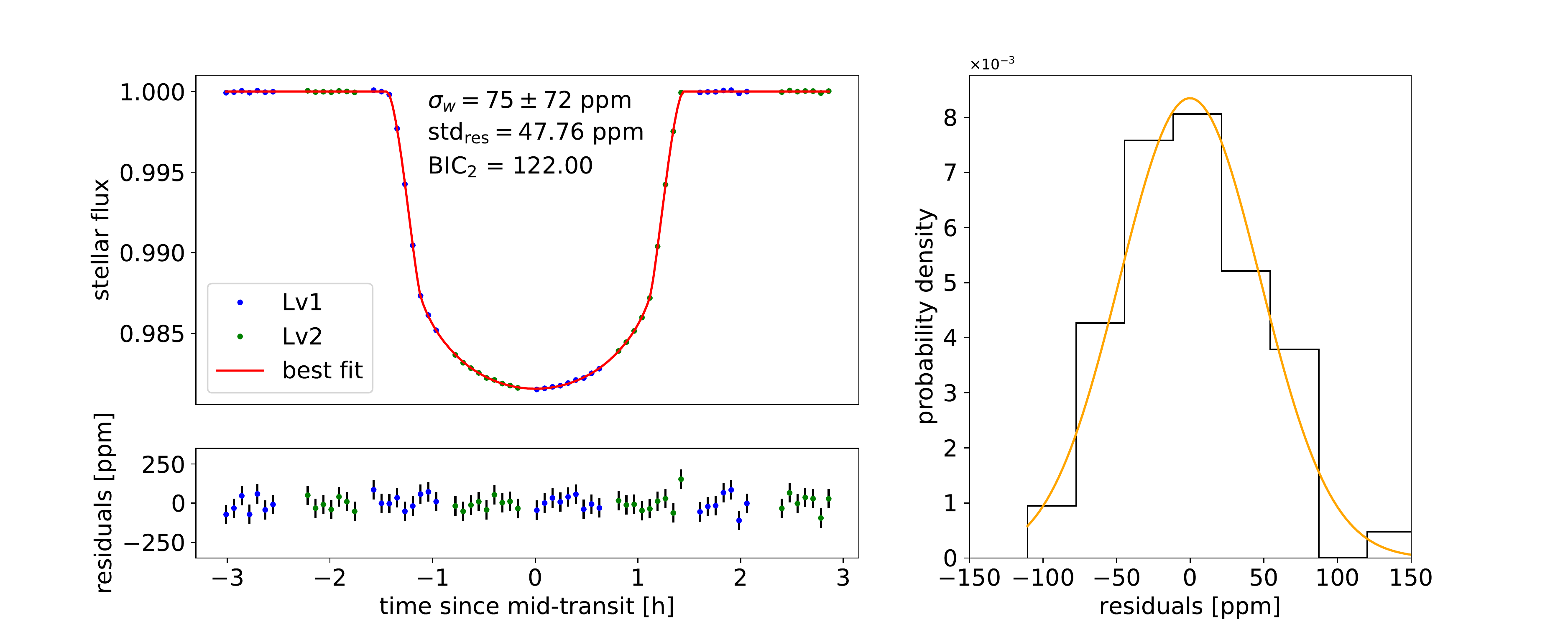}
    \caption{Same as Figure \ref{fig:g430l_model1} but for Model 2.}
    \label{fig:g430l_model2}
\end{figure*}

\begin{figure*}
    \plotone{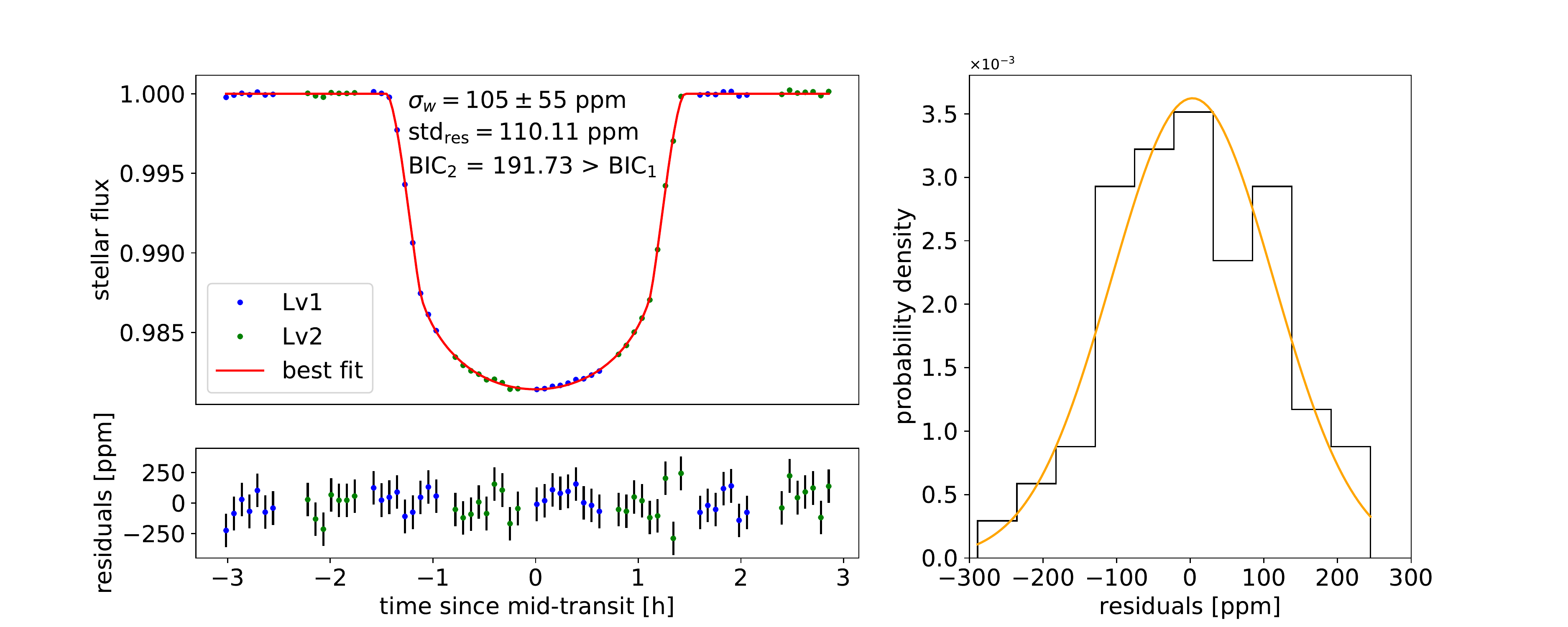}
    \caption{Same as Figure \ref{fig:g430l_model1} but for Model 3.}
    \label{fig:g430l_model3}
\end{figure*}

\begin{figure*}
    \plotone{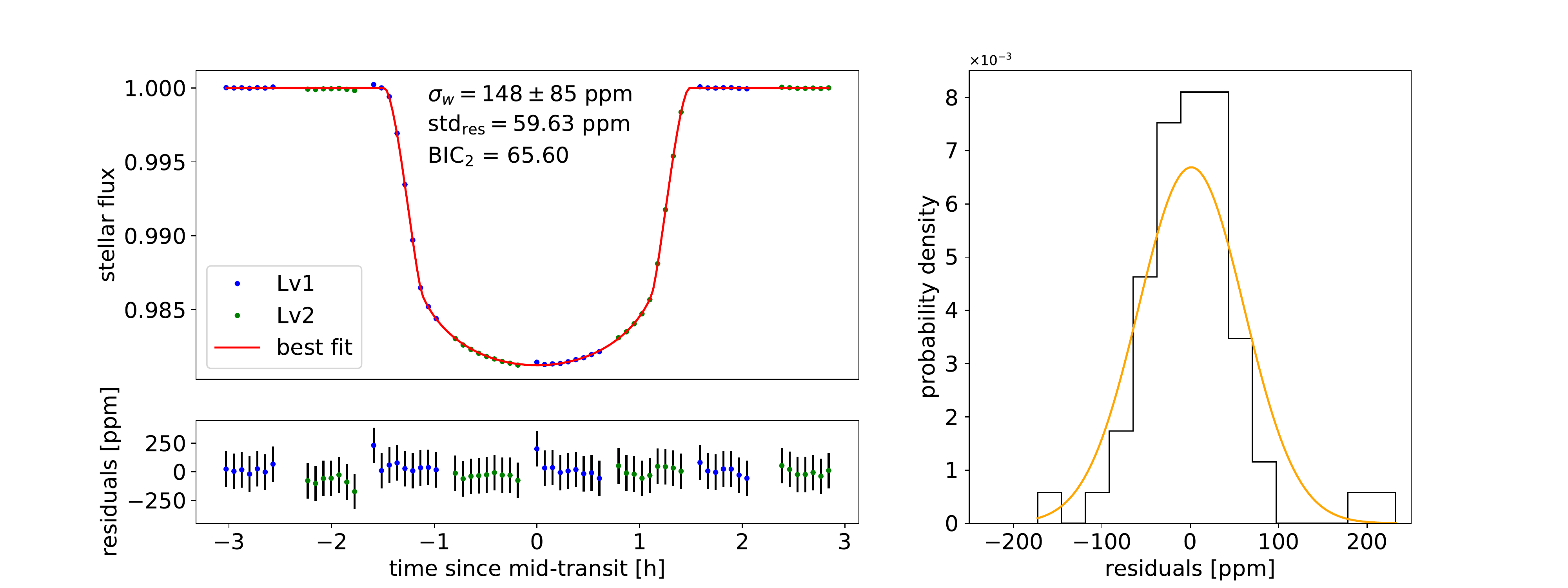}
    \caption{Same as Figure \ref{fig:g430l_model1} but for Model 4.}
    \label{fig:g430l_model4}
\end{figure*}

The noise present in the TLC may be discriminated between uncorrelated noise (white noise) and time-correlated noise or systematics (red noise). The latter may come from stellar flares or instrumental systematics, for instance. The amount of white noise should not be lower than the expected photon noise, $\sigma_{ph}$, for the considered instrument, which can be approximated as \citep{kjeldsen1992}:
\begin{equation}
    \sigma_{\mathrm{ph}} = 0.011D^{-1}\Delta\lambda^{-1/2}\Delta t^{-1/2}10^{0.2V},
\end{equation}
where D is the telescope diameter in centimeter (240 cm for HST), $\Delta\lambda$ is the equivalent width of the filter in nanometer (280 nm for G430L), $\Delta t$ is the exposure time in seconds (253 s), and $V$ is the stellar magnitude in the V band (10.4 for WASP-121). We obtain $\sigma_{\mathrm{ph}} = 21\, \mathrm{ppm/\sqrt{\Delta\,t}}$.\\
\\
There are three major ways of modelling the red noise. If the source and behavior of the systematics are known, the first way consists in fitting a function which best represents that behavior. Examples include a first order polynomial in time, a second order polynomial in the detector position, or a fourth order polynomial in the HST phase. However, the systematics' behavior is often not well - or at least not completely - understood. Therefore the second option consists in modelling the correlated noise through a Gaussian Process (GP). This method requires the selection of some kernel functions, which, in principle, should be physically motivated \citep{foreman2017}. However, there is no clear connection between the thermal breathing of HST and the physical parameters on which we have a handle on. Therefore, we did not investigate the GP method. The last option corresponds to the so-called wavelet method \citep{carter2009}. An in-depth description of the method is beyond the scope of this paper but we summarize here the main idea \citep[see also][]{cubillos2017}. The red noise is assumed to have a power spectral density varying as $1/f^{\gamma}$, where $f$ is the frequency. The residuals are projected onto an orthonornal wavelet basis with two dimensions: scale and location (in time). This transform greatly simplifies the calculation of the likelihood function. The method requires the selection of a wavelet basis, and \citet{wornell1996} and \citet{carter2009} recommend the fourth-order Daubechies wavelet basis, which performs well for $1/f^{\gamma}$ noise. We further followed the recommendation of \citet{carter2009} and fixed $\gamma=1$, which does not decrease the performance of the noise analysis. In summary, the wavelet method is parametrized by only two parameters: $\sigma_w$ and $\sigma_r$ (see footnote \footnote{\citet{carter2009} emphasized that while $\sigma_w$ is the standard deviation of the white noise, $\sigma_r$ is generally not the standard deviation of the correlated noise.}). These parameters are related to the coefficients' variance of the wavelet transform, see \citet{cubillos2017} for additional details.\\
\\
We summarize in Table \ref{tab:models} the models we considered to analyze the TLCs. It is important to note that each coefficient of any polynomial baseline function has to be fitted separately for the Lv1 and Lv2 data sets. Hence, we report the total number of free parameters for the noise models in the Table. Additionally, we always used the wavelet method to calculate the likelihood function for two reasons: (1) it may help taking into account some systematics not well modelled by the baseline functions, and (2) it allows us to consistently compare the models. Therefore, the total number of free parameter for the noise model is equal to 2 (wavelet method) plus twice the number of coefficients of the baseline function (the factor 2 comes from the fact that we corrected the baseline for each visit separately).
\startlongtable
\begin{deluxetable*}{c|cc|c}
\tablecaption{Transit functions considered for the analysis of the light curves. In the Table, $c_0$ is a constant and $p(x^N)$ denotes a N-order polynomial (without the zero-order term) in time ($x=t$), Hubble Space Telescope phase ($x=\phi$), and detector position ($x=X$ and/or $x=Y$). The symbol $tr$ denotes a noise-free transit curve, while $w$ denotes the red noise component estimated from the wavelet method. $N_{free}$ is the number of free parameters for the noise models. \label{tab:models}}
\tablehead{
\colhead{Model \#} & \colhead{Baseline function} & \colhead{Transit function} & \colhead{$N_{free}$}
}
\startdata
model 1 & $B_1 = c_0 + p(t^1)$ \tablenotemark{a} & $tr \cdot B_1 + w$ & 2$\times$2 + 2 \\
model 2 & $B_2 = c_0 + p(t^1 + \phi^4 + X^1 + Y^1)$ \tablenotemark{b} & $tr \cdot B_2 + w$ & 2$\times$8 + 2 \\
model 3 & $B_3 = c_0 \cdot (1 + p(\phi^4)) \cdot (1 + p(X^1)) \cdot (1 + p(Y^1))$ \tablenotemark{c} & $tr \cdot B_3 + w$ & 2$\times$7 + 2 \\
model 4 & $B_4 = 1$ & $tr \cdot B_4 + w$ & 2
\enddata
\tablenotetext{a}{\citep{evans2018}}
\tablenotetext{b}{\citep{alexoudi2018}}
\tablenotetext{c}{\citep{wakeford2018}}
\end{deluxetable*}
We used a Differential Evolution Markov Chain (DE-MC) to explore the parameter space, and we minimize the wavelet-based likelihood function to obtain the posterior distributions of all fitted parameters. We used the Python library MC$^3$ \citep{cubillos2017} with 12 parallel chains, each consisting of roughly 170,000 steps (2 million steps in total). We applied a total burn-in period of 140,000 steps, and checked that convergence was reached by ensuring a Gelman-Rubin test value smaller than 1.01 for every parameter \citep{gelman1992}.\\
\\
As mentioned in \citet{akinsanmi2019} and \citet{hellard2019}, there is a strong correlation between the planetary mean radius and $h_2$, and we have no prior information on these parameters. One way to refine the Love number estimation is to perform a two-step analysis: the first step uses a DE-MC with uniform priors on the planetary mean radius and $h_2$, until a stationary distribution is reached. The second step uses a DE-MC where the prior on the LDCs and planetary mean radius are the posterior distributions from the first step. The final distributions are used for parameter estimations. This two-step analysis allows the DE-MC to be more efficient, and provides refined parameter estimation. It was first suggested by \citet{dunkley2005} and was used in various studies later on \citep[e.g.,][]{umetsu2018,shin2019}. A careful check of the Bayesian Information Criterion (BIC, see Equation \ref{eq:bic}), distribution of the residuals, and fitted white noise value are carried out to ensure that the refinement step best explains the data. We summarize the change in priors in Table \ref{tab:priors}, while the BIC is defined as \citep{kass1995}:
\begin{equation}
    \label{eq:bic}
    \mathrm{BIC} = \chi^2 + \mathrm{k\ln(n)},
\end{equation}
where $\chi^2$ is the usual chi-squared of the fit, k is the number of model parameters, and n the number of data points.\\
\\
The following results are always the outputs from the second step and we emphasize, if relevant, when the refinement step decreases the fit quality (i.e., when $BIC_2 > BIC_1$). We present in Figures \ref{fig:g430l_model1} - \ref{fig:g430l_model4} the systematics-free data, best fit, residuals, and distribution of the residuals for each models (see Table \ref{tab:models}).
\begin{deluxetable}{c|cc}
\tablecaption{Summary of the priors in the two-step analysis. In the Table, post. dist. denotes posterior distribution. The parameters not mentioned in this Table have unchanged, uniform priors. We refer the reader to Section \ref{sec:ld_analysis} for an extensive discussion on LD. \label{tab:priors}}
\tablehead{
\colhead{Parameter} & \colhead{Step 1} & \colhead{Step 2}
}
\startdata
$u_p = u_1 + u_2$ & $\mathcal{N}$(0.6, 0.1) \tablenotemark{a} & post. dist. step 1 \\
$u_m = u_1 - u_2$ & $\mathcal{N}$(0.4, 0.1) \tablenotemark{a} & post. dist. step 1 \\
$R_p/R_s$ & U(0.05, 0.2) & post. dist. fit 1 \\
$h_2$ & U(0, 2.5) & U(0, 2.5)
\enddata
\tablenotetext{a}{\citep{evans2018}}
\end{deluxetable}
\\
\\
\textbf{Model 1:} It is clear from Figure \ref{fig:g430l_model1} that some periodic systematics remain in the residuals. Furthermore, the distribution of the residuals is far from being Gaussian, and $BIC_2 > BIC_1$. Therefore, the first-order polynomial in time does not properly describe the noise behavior in this data set.\\
\textbf{Model 2:} From Figure \ref{fig:g430l_model2} the residuals do not exhibit any periodic behavior anymore, but somehow the in-transit phases are better fitted than the out-of-transit phases. Furthermore, the distribution of the residuals is still not Gaussian. Overall, Model 2 describes better the noise behavior than Model 1, and this is reflected in a lower BIC despite the higher number of parameters.\\
\textbf{Model 3:} From Figure \ref{fig:g430l_model3} it is clear that the distribution of the residuals is not Gaussian, and that the best fit does not describe well the data. Additionally we have $BIC_2 > BIC_1$. Model 3 does not properly describe the noise behavior in this data set.\\
\textbf{Model 4:} This assumption-free model is the best, as shown in Figure \ref{fig:g430l_model4}. There is no periodic behavior in the residuals, and the distribution of the residuals is well approximated by a Gaussian. Two outliers appear in the residual time series, but an analysis without these two points did not lead to a significantly better result. Most importantly, the BIC is roughly twice as low as Model 2, because the noise model contains only two parameters.\\
\\
We conclude from this analysis that the wavelet method alone best describes the noise behavior in the TLC, without overfitting the data. With only two parameters, it prevents the fitting of complicated baseline models with multiple parameters. This is of particular interest when the noise sources are not well - or not fully - understood, as it is often the case.\\
\\
In Table \ref{tab:summary_noise} we summarize the derived BIC and Love number values for each noise model. The errorbars present similar amplitudes, regardless of the model. Although all peak values are within their 1-sigma uncertainty, their differences range from 0.04 (between models 1 and 4) to 0.44 (between models 2 and 4), which would drastically change the interpretation of the planetary interior. Surprisingly, the second best model (model 2) presents the less consistent peak value with the best model (model 4); and a peak value lower than 1 suggesting non-linear effects taking place in the interior. Therefore, the impact of the noise model in the estimation of $h_2$ remains high, though we showed that the wavelet method provides the most reliable result.
\begin{deluxetable}{c|cc}
\tablecaption{Summary of the derived BIC and Love number values, $h_2$, for each noise model.
\label{tab:summary_noise}}
\tablehead{
\colhead{Model \#} & \colhead{BIC} & \colhead{derived $h_2$}
}
\startdata
model 1 & 125.61 & $1.35^{+0.71}_{-0.79}$ \\
model 2 & 122.0 & $0.95^{+0.73}_{-0.60}$ \\
model 3 & 191.73 & $1.31^{+0.70}_{-0.77}$ \\
model 4 & 65.60 & $1.39^{+0.71}_{-0.81}$
\enddata
\end{deluxetable}
\\
\\
In the next subsection we perform a limb darkening study, and show that a careful analysis must be carried out when combining the wavelet method with LD. In particular, we show that the wavelet method can mask limb brightening.

\subsection{Limb darkening study} \label{sec:ld_analysis}
\begin{figure*}
    \plotone{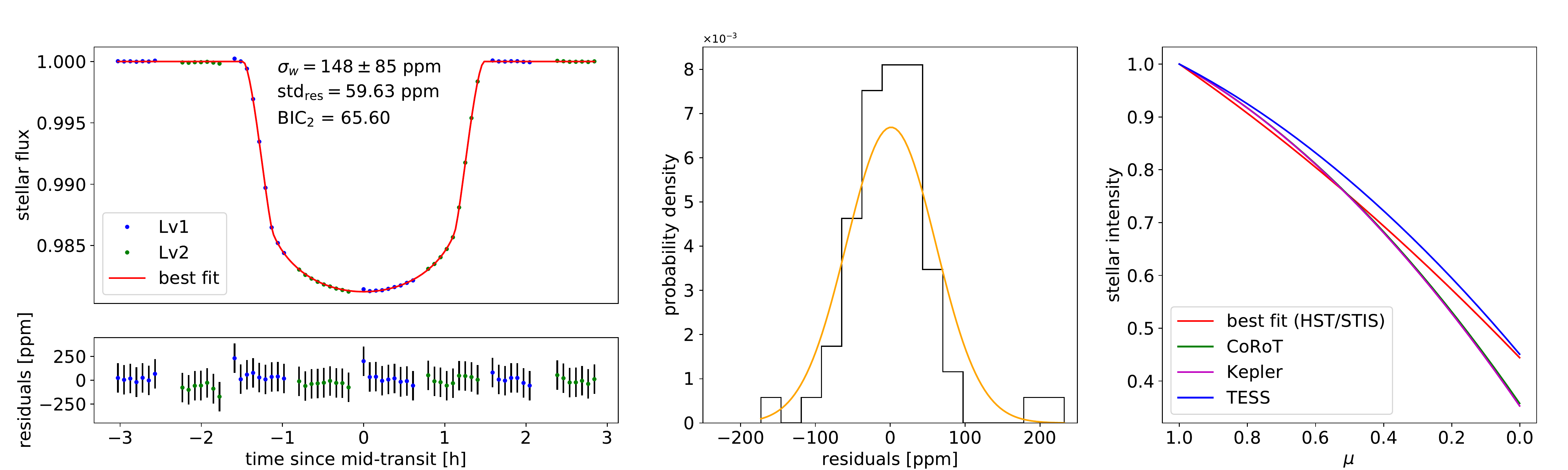}
    \caption{Systematics-free data and best fit (top left panel), residuals with white noise errorbars (bottom left panel), distribution of the residuals (middle panel), and stellar intensity profiles (right panel) for the quadratic LD law. $\mathrm{std_{res}}$ corresponds to the standard deviation of the residuals, while the orange solid line in the middle panel corresponds to the Gaussian fit of the residuals' distribution.}
    \label{fig:g430l_quadLD}
\end{figure*}

\begin{figure*}
    \plotone{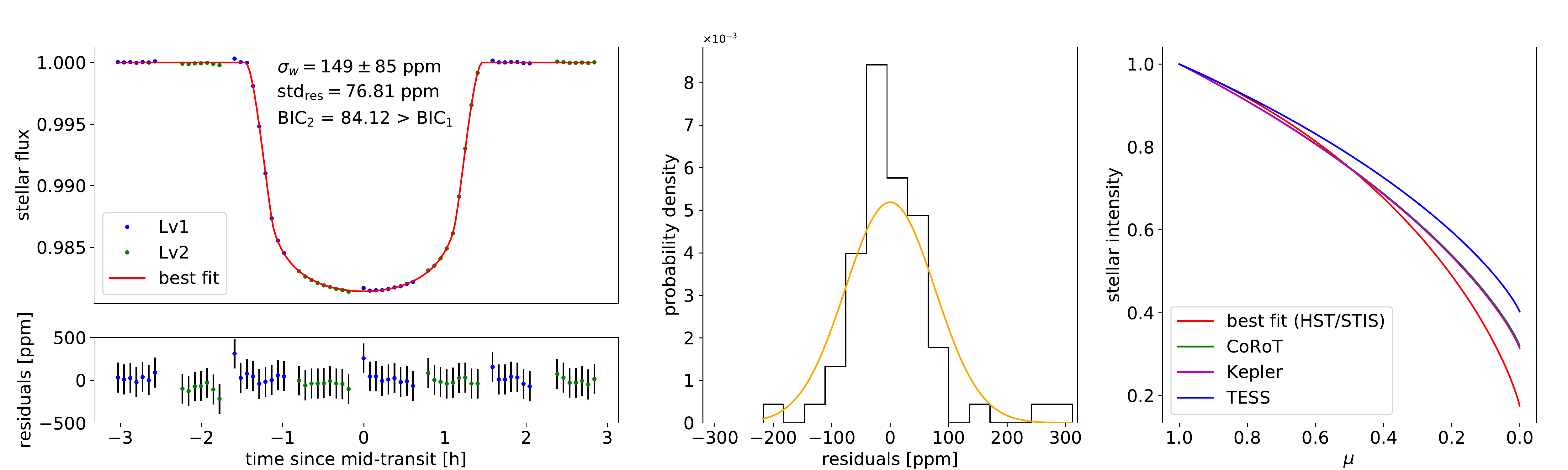}
    \caption{Same as Figure \ref{fig:g430l_quadLD} but for the logarithmic LD law.}
    \label{fig:g430l_logLD}
\end{figure*}

\begin{figure*}
    \plotone{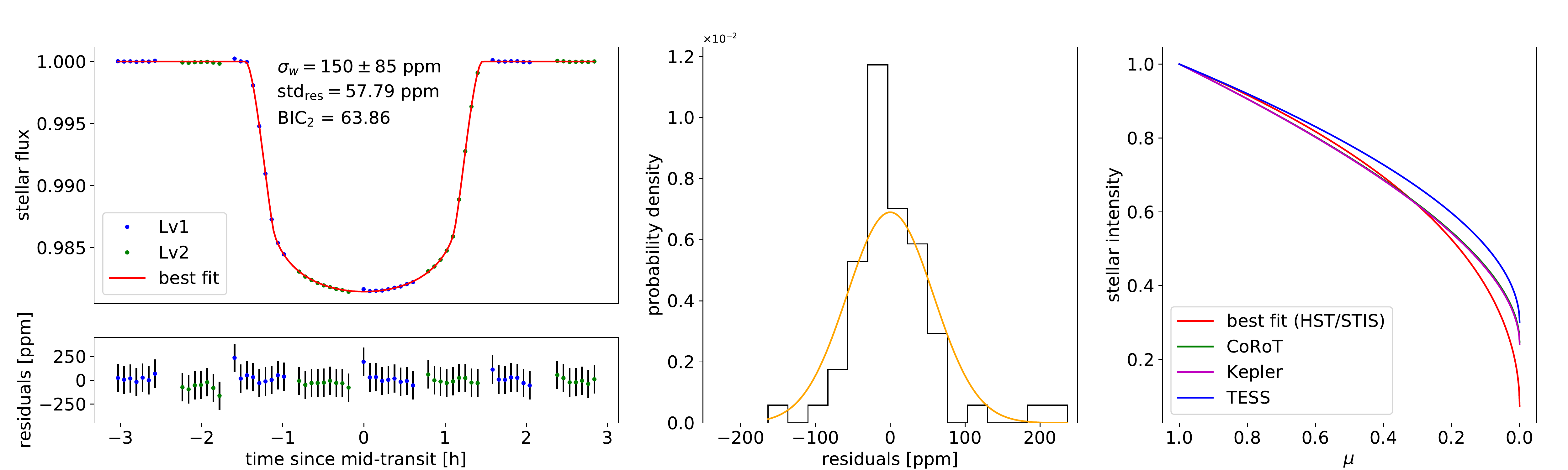}
    \caption{Same as Figure \ref{fig:g430l_quadLD} but for the square-root LD law.}
    \label{fig:g430l_sqrtLD}
\end{figure*}

\begin{figure*}
    \plotone{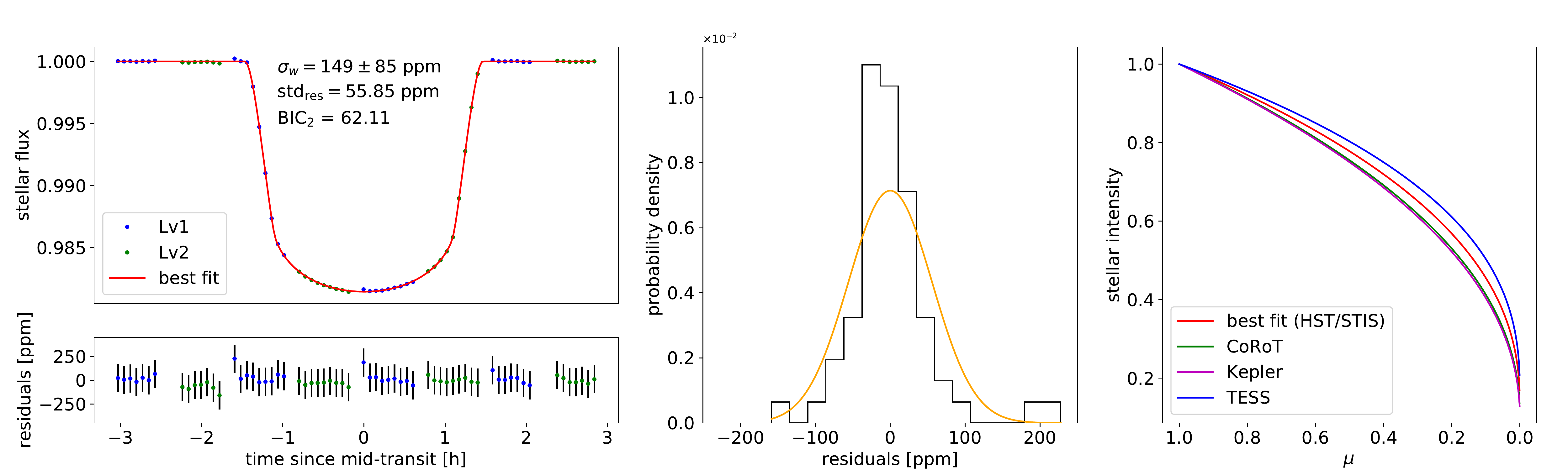}
    \caption{Same as Figure \ref{fig:g430l_quadLD} but for the power-2 LD law.}
    \label{fig:g430l_power2LD}
\end{figure*}

\begin{figure*}
    \plotone{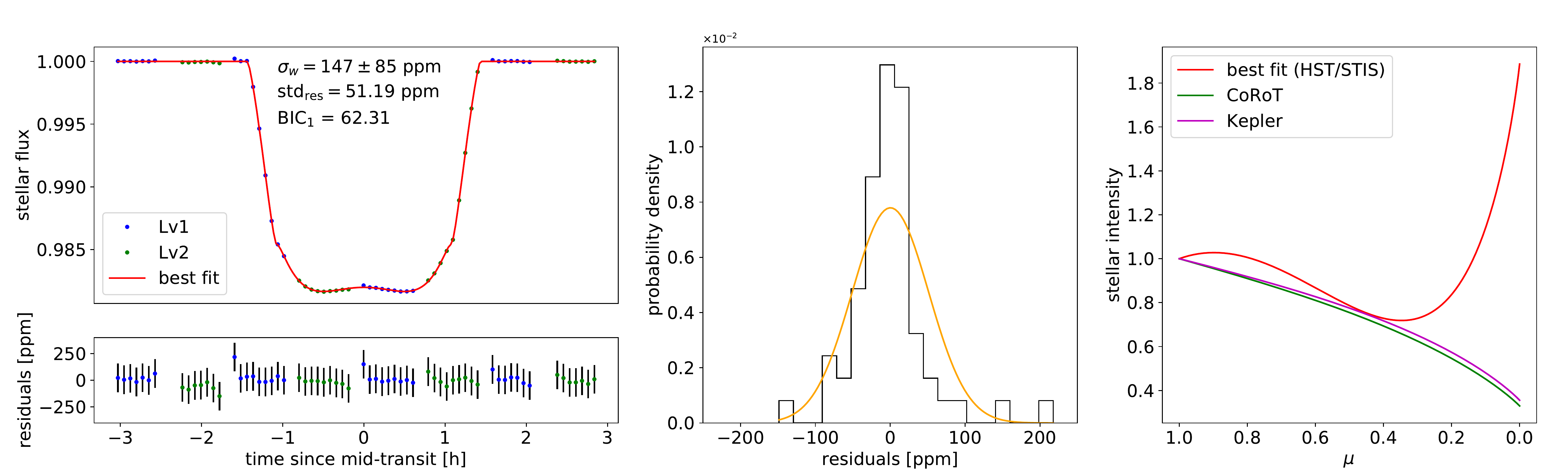}
    \caption{Same as Figure \ref{fig:g430l_quadLD} but for the 3-parameter LD law. Output from step 1.}
    \label{fig:g430l_3paramLD}
\end{figure*}

\begin{figure*}
    \plotone{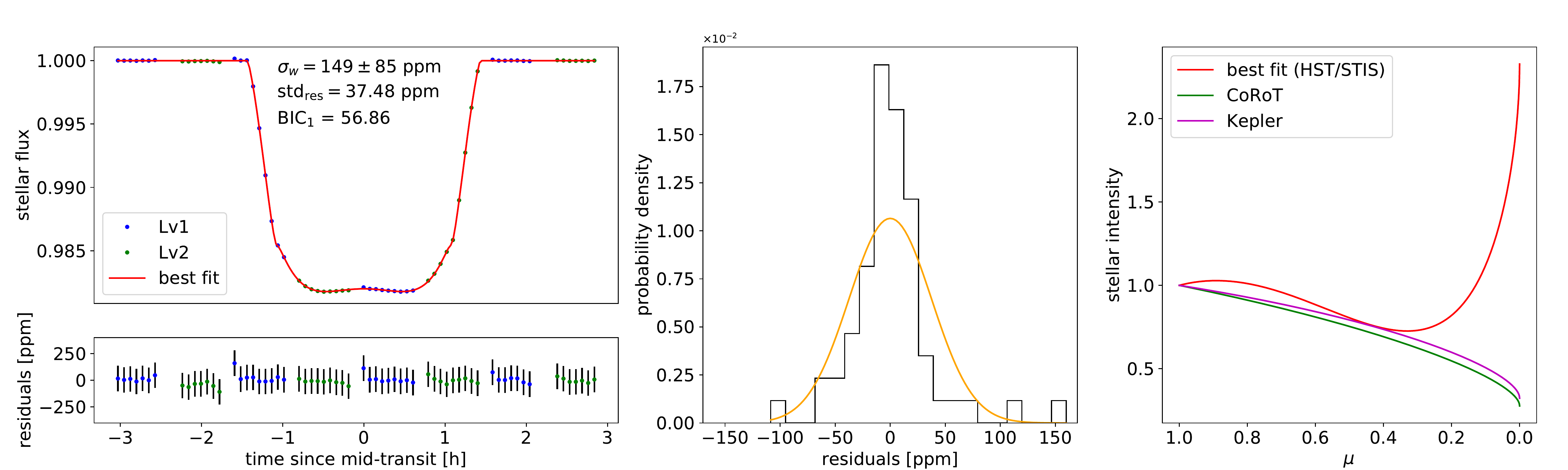}
    \caption{Same as Figure \ref{fig:g430l_quadLD} but for the 4-parameter LD law. Output from step 1.}
    \label{fig:g430l_4paramLD}
\end{figure*}

In this subsection the model used to fit the data is Model 4 from Table \ref{tab:models}. We test several LD laws, in addition to the quadratic one defined as \citep{kopal1950}:
\begin{equation}
    \frac{I(\mu)}{I_0} = 1 - u_1(1-\mu) - u_2(1-\mu)^2,
\end{equation}
where $\mu=\cos(\gamma)$, with $\gamma$ the angle between the direction to the observer and the normal to the stellar surface. $I_0$ is the normal emergent intensity at the stellar center.\\
The logarithmic law is defined as \citep{klinglesmith1970}:
\begin{equation}
    \frac{I(\mu)}{I_0} = 1 - u_1(1-\mu) - u_2\mu\mathrm{ln}(\mu).
\end{equation}
The square-root law is defined as \citep{diaz1992}:
\begin{equation}
    \frac{I(\mu)}{I_0} = 1 - u_1(1-\mu) - u_2(1-\sqrt{\mu}).
\end{equation}
The power-2 law is defined as \citep{hestroffer1997,maxted2018}:
\begin{equation}
    \frac{I(\mu)}{I_0} = 1 - c(1-\mu^{\alpha}).
\end{equation}
The three-parameter law is defined as \citep{sing2009}:
\begin{equation}
    \frac{I(\mu)}{I_0} = 1 - \sum_{i=2}^{i=4}u_i(1-\mu^{i/2}),
\end{equation}
while the four parameter law is defined as \citep{claret2000}:
\begin{equation}
    \frac{I(\mu)}{I_0} = 1 - \sum_{i=1}^{i=4}u_i(1-\mu^{i/2}).
\end{equation}
As described in Section \ref{sec:noise_analysis}, we perform two fits by updating the priors on the LDCs and planetary mean radius. For the quadratic LD law we have estimated values for the LDCs in the G430L waveband from the literature \citep{evans2018}. However, no estimated values for any other LD law are available in this specific waveband and/or for these stellar parameters (i.e., effective temperature and metallicity). From a theoretical point of view, some tables are available in the literature for several LD law, mostly for other wavebands relevant to the CoRoT, Kepler, and TESS instruments. The Python package ldtk \citep{parviainen2015} allows the user to compute theoretical model-specific LDCs using the PHOENIX library \citep{husser2013}. We find that even for the quadratic law, the theoretical LDCs are far from the fitted values. Hence, we did not use any priors from this package. Additionally, it is not clear what the bounds for the LDCs are, especially for the three- and four-parameter laws. Thus one has to ensure the fitted stellar intensity profile is nowhere negative or does not present limb brightening. Although the latter effect can appear in the presence of bright spots or plages \citep{csizmadia2013}, quiet stars are more appropriate for such fine TLCs analyses. Therefore, we wish to avoid limb brightening by target selection.\\
\\
We show in Figures \ref{fig:g430l_quadLD} - \ref{fig:g430l_4paramLD} the results for each LD law, in the same fashion as in Figures \ref{fig:g430l_model1}
- \ref{fig:g430l_model4}, but we add the best-fitted stellar intensity profile and compare it with theoretical expectations for other observing facilities. The latter were chosen based on their wavelength coverage (as consistent as possible with the G430L grating) and on the availability of LDCs tables \citep{king2010,claret2011,claret2017,maxted2018}. For the 3-parameter and 4-parameter laws, only the output from step 1 (see Table \ref{tab:priors}) is shown, see below for details.
\\
\\
\textbf{Quadratic LD:} Same as Model 4 in the previous section. In addition, the stellar intensity profile is in agreement with theoretical calculations for other observing facilities.\\
\textbf{Logarithmic LD:} The standard deviation of the residuals increases and the Gaussianity of its distribution decreases. Furthermore we have $BIC_2 > BIC_1$. Although the best fitted stellar intensity profile agrees with theoretical calculations, we conclude that the stellar LD is not properly modelled by a logarithmic law in this data set.\\
\textbf{Square-root LD:} The standard deviation of the residuals and derived BIC are smaller than the ones obtained with a quadratic LD law. The stellar intensity profile fits to theoretical expectations in other wavebands. However, the Gaussianity of the residual distribution decreases. Although the derived $h_2$ agrees with the quadratic LD law, the best-fit value is at the edge of the 1$\sigma$ confidence interval. Therefore, we do not favour the square-root law.\\
\textbf{power-2 LD:} As for the square-root law, the standard deviation of the residuals and BIC are slightly smaller than the ones obtained with the quadratic LD law. Furthermore, the stellar intensity profile fits the expectations in other wavebands. However, the Gaussianity of the residual distribution slightly decreases. Although the fitted $h_2$ agrees with the quadratic law, its precision somewhat decreases because the LDCs are not well constrained. Consequently we decide not to favour the power-2 law, despite the slight improvement in the fit. We acknowledge that this law may best explain the stellar LD in this data set, only with better constrained LDCs.\\
\textbf{3-parameter LD:} Despite a lower standard deviation of the residuals and BIC value than the ones derived for the quadratic LD law, it is striking that the transit shape is unusual. The first indication of a low-quality model comes from the distribution of the residuals, which happens to be far from Gaussian. The main scientific information indicating a wrong modelling lies in the stellar intensity profile, which exhibits limb brightening close to the stellar center and toward the edge of the stellar disk.\\
\textbf{4-parameter LD:} The exact same comments emphasized for the 3-parameter LD law hold for the 4-parameter law. The standard deviation of the residuals and the BIC value are even smaller. However, the stellar intensity profile shows a limb brightening close to the stellar center and toward the edge of the stellar disk.\\
\\
In Table \ref{tab:summary_ld} we summarize the derived BIC and Love number values for each LD law. The peak values remain consistent with each other, with differences ranging from 0.02 (between the quadratic, logarithmic, and square-root laws) to 0.05 (between the logarithmic, square-root, and power-2 laws). The peak value of the second best law (power-2, 1.36) is only lower than the one of the best law (quadratic, 1.39) by 0.03. Therefore, it appears that the impact of the limb-darkening modelling in the determination of $h_2$ is weaker compared to the impact of the systematics' modelling (see Section \ref{sec:noise_analysis}).
\begin{deluxetable}{c|cc}
\tablecaption{Summary of the derived BIC and Love number values, $h_2$, for each LD law.
\label{tab:summary_ld}}
\tablehead{
\colhead{LD law} & \colhead{BIC} & \colhead{derived $h_2$}
}
\startdata
quadratic & 65.60 & $1.39^{+0.71}_{-0.81}$ \\
logarithmic & 84.12 & $1.41^{+0.71}_{-0.82}$ \\
square-root & 63.86 & $1.41^{+0.71}_{-0.82}$ \\
power-2 & 62.11 & $1.36^{+0.73}_{-0.82}$ \\
3-parameter & \textit{62.31} \tablenotemark{a} & --- \\
4-parameter & \textit{56.86} \tablenotemark{a} & ---
\enddata
\tablenotetext{a}{BIC output from step 1. These LD laws present nonphysical limb brightening, therefore we do not report the derived Love number.}
\end{deluxetable}
\\
\\
We showed that in the presence of poorly constrained LDCs, one has to be very cautious when using the wavelet method for correlated noise analysis. This method may mask limb brightening while still providing a reasonable light curve. Hence, one must verify the distribution of the residuals, and most importantly, the fitted stellar intensity profile. We found that a quadratic law best describes the stellar LD in this data set. This is in agreement with most transit light curve analyses, since quadratic LDCs are the most studied, hence the most constrained. The community should strive to improve our understanding of stellar LD as it plays a crucial role in transit curve analysis in general, and in the precision of Love number measurements in particular.

\section{Discussion} \label{sec:discussions}
\begin{figure*}
\plotone{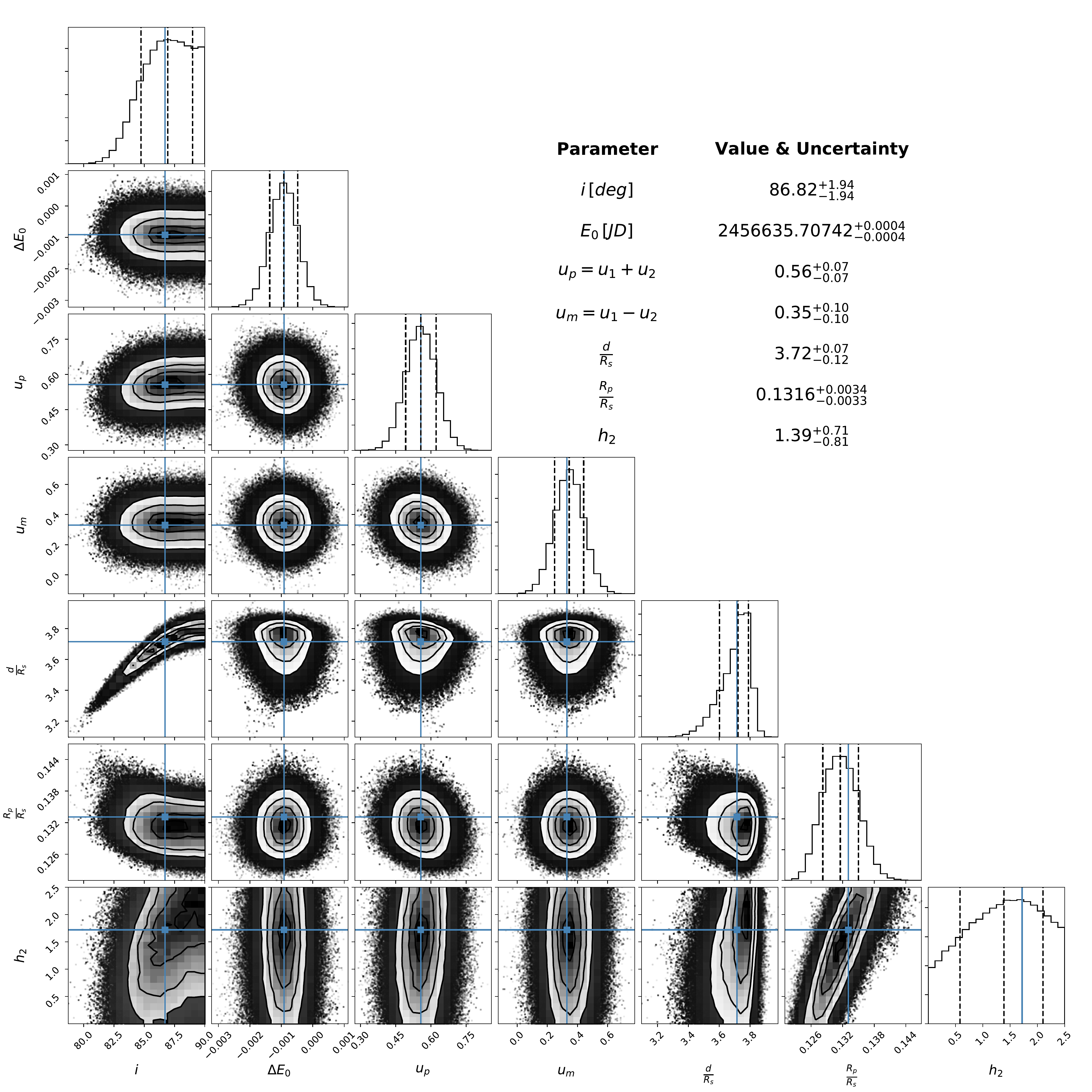}
\caption{Posterior distributions and correlation plots of the transit parameters (noise Model 4 and quadratic LD law, see Section \ref{sec:lc_analysis}). Dotted lines represent the mean and 68\% interval confidence, while the plain blue line corresponds to the best fit.}
\label{fig:g430l_model4_posterior}
\end{figure*}

The analysis presented in the previous Section leads to a tentative 2-sigma detection of the Love number, $h_2 = 1.39^{+0.71}_{-0.81}$, and the posterior distributions of the transit parameters are summarized in Figure \ref{fig:g430l_model4_posterior}. Even though the uncertainty is high, it is essential to note that the peak of the distribution is far from 0, a value corresponding to a spherical body. Additionally, the peak is not far from the latest Jovian measurement of 1.62 \citep{iess2018}, and it matches the measurement for Saturn of 1.39 \citep{lainey2017}. We performed two additional analyses where we either set 1.0 as $h_2$'s lower bound (as the interior of the hot Jupiter WASP-121b is expected to be in hydrostatic equilibrium) or extended the bounds on $h_2$ to $0 < h_2 < 3.5$ (to accommodate unexpected distortions) but found no improvement in the estimation of $h_2$.\\
\\
\citet{hellard2019} showed that a maximal white noise level of 90 ppm/$\sqrt{\mathrm{1\,minute}}$ is required to reliably retrieve $h_2$. We calculated a white noise level of $\sigma_w = 148 \pm 85\, \mathrm{ppm/\sqrt{\Delta t}}$ for one complete observed transit. Assuming the systematics can be removed as showed in Section \ref{sec:noise_analysis}, the white noise level reached after N complete observed transits, per minute exposure, $\sigma_N$, is given by:
\begin{equation}
    \label{eq:n_transits}
    \sigma_N = \sigma_w\sqrt{\frac{\Delta t}{1\,\mathrm{minute}}}\sqrt{\frac{1}{N}}.
\end{equation}
Using $\sigma_w=148 \mathrm{ppm/\sqrt{\Delta t}}$ and Equation (\ref{eq:n_transits}), we deduce that 12 complete observed transits are needed to reach 90 ppm/$\sqrt{\mathrm{1\,minute}}$. While this is valid for the G430L grating, the required number of observed transits could decrease by using a wider wavelength coverage (e.g., the G141 grism of the Wield Field Camera 3). This proves that the Hubble Space Telescope has the capability to carry out Love number measurements. 
\\
\\
Theoretical models for homogeneous H-He planets indicate that an object with the mass of WASP-121b would have a radius similar to Jupiter \citep[e.g.,][]{Fortney2007}. The measured $1.9$ ${\rm R_{Jup}}$ indicates beyond doubt that WASP-121b is an inflated hot-Jupiter. Several mechanisms have been proposed to provide a suitable energy source for the inflation of hot-Jupiters, with Ohmic dissipation \citep{Batygin2010} being possibly the most likely \citep{Thorngren2018}. The interior density profile of a non-inflated planet with the mass of WASP 121b would be relatively well approximated by a polytropic relation of index $n=1$ \citep{Guillot2005}, which has a fluid Love number $h_2$ of about 1.52 \citep[using the Matrix-propagator approach, see][]{padovan2018}. The lower central value of 1.39 inferred in this study provides a further confirmation of the inflated nature of WASP-121b. Given the dependence of $h_2$ on 
the density profile in the interior \citep[e.g.,][]{padovan2018}, and the different locations of the energy deposition for different inflation mechanisms \citep[e.g.,][]{Komacek2017}, improving the error bars on future determinations of $h_2$ of hot-Jupiters, using the method proposed here, will provide a direct way to better constrain the physics of inflation.

\section{Conclusions} \label{sec:conclusion}
Using two primary transits of WASP-121b precisely observed by the Imaging Spectrograph on-board the Hubble Space Telescope, we tested the instrument capability for Love number measurements. We first performed a thorough noise analysis and showed that the wavelet method alone best describes the time-correlated behaviour of the noise, without the need to fit complicated baseline functions with many parameters. Second, we performed a stellar limb darkening study and found that it is best described by a quadratic law. Most importantly we emphasized that, in some cases, the wavelet method can mask limb brightening when the limb darkening coefficients are poorly constrained. Therefore we strongly encourage the community to further study stellar limb darkening processes, as it plays a crucial role in fine transit modelling. Our analysis provided a tentative 2-sigma detection of the Love number: $h_2 = 1.39^{+0.71}_{-0.81}$, in which the impact of the noise modelling in its determination was found to be stronger than the impact of the limb darkening modelling. We showed that a total of 12 complete observed transits are required to precisely estimate $h_2$ with the G430L grating. This number could decrease by increasing the wavelength coverage, e.g., using the G141 grism of the Wide Field Camera 3. A precise estimate of $h_2$ would refine the constraints on the interior of WASP-121b and provide direct insights on the physics of inflation.

\acknowledgments
We acknowledge support from the DFG via the Research Unit FOR 2440 \textit{Matter under planetary interior conditions}. Sz.Cs. thanks the Hungarian National Research, Development and Innovation Office for the NKFI-KH-130372 grants. We warmly thank Thomas M. Evans for kindly sharing his Hubble Space Telescope observations of WASP-121b.



\end{document}